\documentclass[sigconf]{acmart}

\usepackage{booktabs}
\usepackage{tabularx}
\usepackage{multirow}
\usepackage{graphicx}
\usepackage{amsmath}
\usepackage{calc}

\newcommand{\system}{GestureMap}
\newcommand{\systems}{GestureMap's}

\definecolor{changeNoteColor}{rgb}{0.1,0.6,1}
\newcommand{\changenote}[1]{#1} %

\AtBeginDocument{%
  \providecommand\BibTeX{{%
    \normalfont B\kern-0.5em{\scshape i\kern-0.25em b}\kern-0.8em\TeX}}}

\copyrightyear{2021}
\acmYear{2021}
\setcopyright{acmlicensed}\acmConference[CHI '21]{CHI Conference on Human Factors in Computing Systems}{May 8--13, 2021}{Yokohama, Japan} \acmBooktitle{CHI Conference on Human Factors in Computing Systems (CHI '21), May 8--13, 2021, Yokohama, Japan}
\acmPrice{15.00}
\acmDOI{10.1145/3411764.3445765}
\acmISBN{978-1-4503-8096-6/21/05}

\begin{document}

\title{\system: Supporting Visual Analytics and Quantitative Analysis of Motion Elicitation Data by Learning 2D Embeddings}
\renewcommand{\shorttitle}{\system: Supporting Visual Analytics and Quantitative Analysis of Motion Elicitation Data}

\author{Duong Hai Dang}
\email{duong.dang@uni-bayreuth.de}
\author{Daniel Buschek}
\orcid{0000-0002-0013-715X}
\email{daniel.buschek@uni-bayreuth.de}
\affiliation{%
  \institution{Research Group HCI + AI, Department of Computer Science, University of Bayreuth}
  \city{Bayreuth}
  \state{Germany}
}

\renewcommand{\shortauthors}{Dang and Buschek}

\begin{abstract}

This paper presents \textit{\system}, a visual analytics tool for gesture elicitation which directly visualises the space of gestures. Concretely, a Variational Autoencoder embeds gestures recorded as 3D skeletons on an interactive 2D map.
\textit{\system} further integrates three computational capabilities to connect exploration to quantitative measures: Leveraging DTW Barycenter Averaging (DBA), we compute average gestures to 1) represent gesture groups at a glance; 2) compute a new consensus measure (variance around average gesture); and 3) cluster gestures with k-means.
We evaluate \textit{\system} and its concepts with eight experts and an in-depth analysis of published data. Our findings show how \textit{\system} facilitates exploring large datasets and helps researchers to gain a visual understanding of elicited gesture spaces. It further opens new directions, such as comparing elicitations across studies.
We discuss implications for elicitation studies and research, and opportunities to extend our approach to additional tasks in gesture elicitation.

\end{abstract}



\begin{teaserfigure}
  \includegraphics[width=\linewidth]{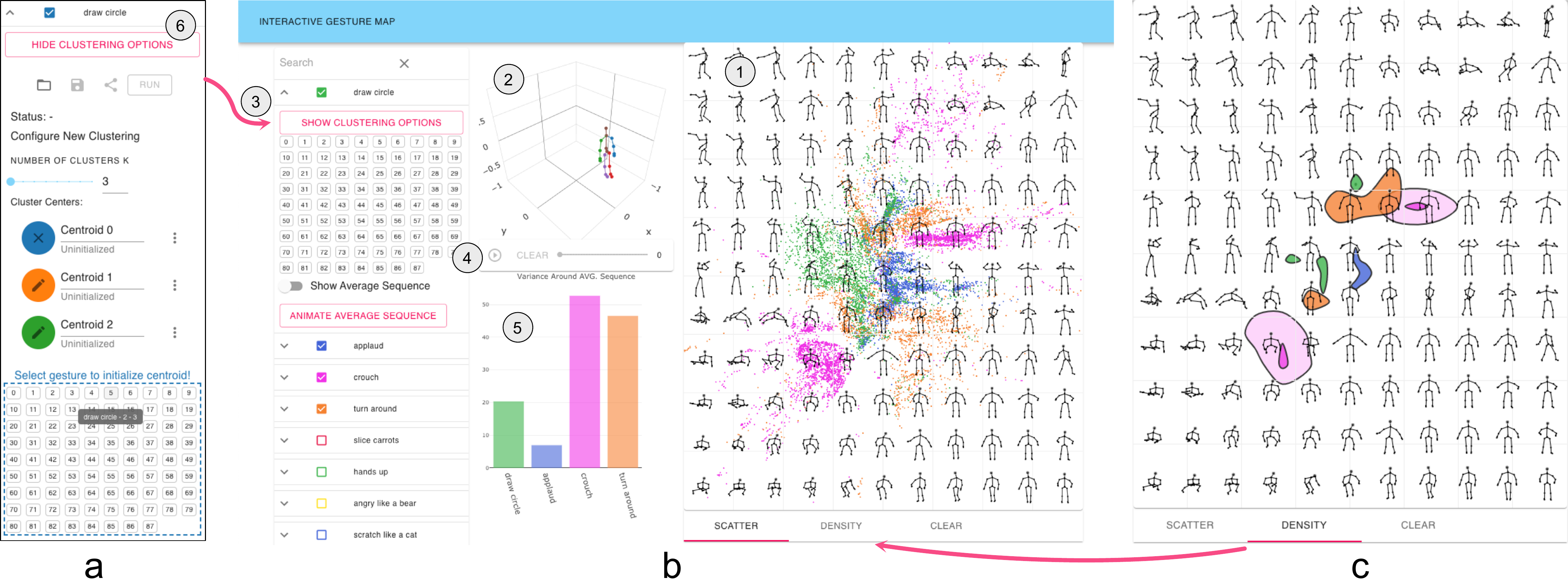}
  \caption{Our visual analytics tool, \textit{\system}: A configuration view to initialize and run the interactive clustering algorithm (a), an overview of the entire application window (b), a density plot projected onto the gesture map (c). The numbers mark different components further explained in Section~\ref{sec:tool}.}
  \Description{Three modules from the application are shown side-by-side. The different components are further described in the text.}
  \label{fig:teaser}
\end{teaserfigure}

\maketitle

\section{Introduction}

Designing effective interactions and user interfaces often involves exploring two potentially high-dimensional spaces~\cite{Williamson2012}: 1) The space of human behaviour (e.g. comfortable motion ranges of arm and hand), and 2) the space of senseable input in a system or context (e.g. tracking of up to X body joints in 3D).

Although central to HCI, the field has developed few dedicated methods and tools for supporting the (joint) exploration of such user-sensor spaces (cf.~\cite{Williamson2012}). 
One successful method that has seen widespread use is the elicitation study paradigm~\cite{Wobbrock2005}, which helps HCI researchers and practitioners to explore the space of possible and ``intuitive'' or ``guessable'' (gesture) commands:
Participants are shown a ``referent'' (often a system action, e.g. \textit{volume up}) and are asked to propose and perform a gesture they would use for it (e.g. \textit{turn wrist right}). This is repeated for several referents.

Researchers then analyse these gesture proposals, compute measures to identify common proposals (e.g.~\cite{Vatavu2019, Wobbrock2005}), and decide on a set of gestures to be used in an interactive system, typically composed of the gestures with high agreement among participants (e.g.~\cite{Vatavu2012, Wobbrock2009}).

In this way, elicitation studies inform gestural interaction with user-driven exploration: Most studies focus on the \textit{human behaviour space} and thus do not rely on a specific sensor; they typically video-record participants for manual gesture analysis (e.g.~\cite{E2017, Kim2020}). Some additionally employ a sensor in elicitation (e.g. Leap~\cite{Vatavu2014}, Kinect~\cite{Vatavu2012}), thus also potentially considering the \textit{senseable space}.%

While elicitation studies have become a widely used staple in the HCI toolbox, they still present challenges (cf.~\cite{Tsandilas2018, Villarreal-Narvaez2020}), including the need for \textit{manual data analysis}. This limits elicitation studies, as well as the general endeavor of systematically exploring behaviour-sensor spaces in HCI, as characterised in the following paragraphs:

\begin{itemize}
    \item \textit{Workload:} Watching videos to manually classify gestures (e.g.~\cite{E2017, Kim2020, Vatavu2012, Vatavu2014, Wobbrock2009}) is tedious work~\cite{Tsandilas2018}. It may thus also hinder the use of elicitation in user-centred processes that require repeating such work (e.g. iteration).
    \item \textit{Subjectivity:} As critically pointed out~\cite{Tsandilas2018}, manual interpretation at best requires further efforts (e.g. multiple coders); at worst, it leads to subjectively biased results.
    \item \textit{Limited scale:} Without quantitative analysis tools, large-scale elicitation remains scarce (survey: mean=25 participants~\cite{Villarreal-Narvaez2020}). This stands in contrast to motivations for diverse samples and for training  recognisers on elicited gestures~\cite{E2017, Kane2011, Smith2018, Villarreal-Narvaez2020}. 
    \item \textit{Isolated results:} The lack of quantitative data analysis methods and tools hinders replication, reuse, extensions, and comparisons across elicitations, even if the same sensor was used (e.g. Kinect). Most work collects and analyses new data~\cite{Villarreal-Narvaez2020}, isolated from previously published datasets.
\end{itemize}

These challenges motivate our work on new quantitative methods and tools for analyzing elicitation data. Fittingly, recent related work highlighted the need and feasibility of more objective, computational measures~\cite{Vatavu2019}, and called for further computational models and measures, based on a survey of 216 elicitation studies~\cite{Villarreal-Narvaez2020}.

Addressing this, we extend the computational toolbox for analyzing gesture elicitation data with these contributions:

\begin{itemize}
    \item \textit{\system}, a method and tool for visualizing and exploring motion data from elicitation studies on an interactive, learned 2D map, inspired by concepts from visual analytics.
    \item \textit{New computational capabilities} for gesture representation, consensus and clustering, based on average gestures computed with DTW Barycenter Averaging (DBA)~\cite{Petitjean2011}, to connect exploration to quantitative measures, extending the computational approach motivated in recent work~\cite{Vatavu2019, Villarreal-Narvaez2020}.
    \item \textit{Insights} from using \textit{\system} in a detailed case study on datasets from the literature, plus a qualitative expert evaluation with eight researchers.
\end{itemize}

\section{Related Work}
The analysis concepts introduced in this work are built on previous work spanning HCI, machine learning and visual analytics. This section briefly describes gesture elicitation studies, followed by an overview of tools that support researchers across different tasks involved in analyzing elicitation data. In particular, we outline existing analysis concepts for high-dimensional data.

\subsection{Gesture Elicitation Studies}

The gesture elicitation paradigm was first introduced by Wobbrock et al. \cite{Wobbrock2005} to elicit users' interaction preferences for new systems. This method was then specifically adapted to include gesture proposals to control surface tabletop computers \cite{Wobbrock2009}. In a subsequent study, Morris et al. \cite{Morris2009} confirmed that new users do prefer the user-defined gesture set over the one created by experts. Since then, this method has become a standard tool for the design of gesture input mappings for new interactive systems, for example to control a swarm of robots \cite{Kim2020}, smart-home appliances~\cite{Vatavu2012, Kuehnel2011}, or AR/VR applications \cite{Piumsomboon2013}.

Central to gesture elicitation studies is an in-depth analysis of the proposed data to find common behavior. Researchers have therefore developed various measures to formalize the consensus among participants \cite{Wobbrock2009, Vatavu2012, Vatavu2014, Vatavu2016, Vatavu2013}.

However, these measures rely on subjectively assessing the similarity of the observed gestures: They require researchers to group proposals into subgroups that they consider identical, which is usually done by manual annotation based on watching videos of the participants in the study \cite{Morris2012, Kim2020}. Thus, while these measures set standards on how to compute consensus from gesture proposal, they cannot avoid subjectivity per se.

To address this,  \citet{Vatavu2019} has recently proposed a new, data-driven approach: It employs a distance measure as an objective basis for assessing consensus in elicitation studies. Our work builds on this idea, extends its data-driven perspective with a visual analytics tool, and introduces a new measure fitting this visualization.

\subsection{Gesture Analysis Tools}
Several tools have been created for more effective and objective analyses. Video analysis has been the preferred evaluation method, but the annotation of individual video sequences can be time-consuming~\cite{Tsandilas2018}. An efficient analysis becomes even more important as large-scale gesture data sets can be collected online, for example, through cloud elicitation tools \cite{Ali2019}. Thus, researchers devised different ways to distribute the work among people \cite{Magrofuoco2019, Ali2018}.

While the concepts introduced in this paper also enable researchers to better annotate sequences, our focus lies in particular on the \textit{exploration} of elicited gesture data.

\citet{Nebeling2015} created a tool to analyze recordings created by a Kinect camera sensor. They included three visualizations. First, they used a 3D animation of a Kinect skeleton. Second, they provided a visualization where only the moving joint is drawn on the canvas. The third visualization is similar to the second, but additionally employs a heat-map to emphasize the time domain.

The most similar work to ours is \textit{GestureAnalyzer} by \citet{Jang2014} which also focuses on the analysis of gesture elicitation studies. \changenote{To find behavioral patterns, it employs a variation of the small-multiples plot \cite{Tufte1986} and an interactive hierarchical clustering interface visualized in a tree layout. Their calculations and analyses are based on hand-engineered features. The gesture map which we propose in this work facilitates a richer exploration of the behavior space using machine learned features for the gesture poses. It provides an overview of the gesture data and introduces a new continuous traceable 2D paths which represent gesture sequences. For further discussions on the comparison of these two systems we refer to section \ref{sec:diss:clustering}.}

\subsection{Visualization of High Dimensional Data}
A key challenge in visual analytics is the effective visualization of high-dimensional data. This typically involves two steps: 1) Projecting the data to 2D for display on a screen. 2) Suitably visualising the projected data, considering the analysts' tasks and goals.
While there exist many dimension reduction techniques \cite{Maaten2009, Van2012, McInnes2018, Goodfellow2014, Lawrence2003}, we use a Variational Autoencoder \cite{Kingma2013} to reduce the dimensions of the raw sensor data. To visualize temporal data, a common representation is a line plot, horizon plot \cite{Few2008}, or a small-multiples plot \cite{Tufte1986}. However, these highly abstract visualizations may occlude the nature of the underlying data. For example, these plots may hide the structure of a 3D skeleton recording. We therefore combine an abstract 2D mapping with a grid of representative 3D skeletons to give analysts a visual overview of the proposed gestures.

Also related to our work are tools to analyze and visualize machine learned representations of complex data: Deep learning models are capable of learning human-understandable features of high-dimensional data: For example, \citet{Kingma2013} and \citet{Lawrence2004} sample multiple points from the learned space and visualize them to demonstrate that the learned space is continuous and smooth, but without providing interaction functionalities. Smilkov et al. \cite{Smilkov2016} filled this gap by providing a generic tool to visualize these embeddings.

Some researchers created specific visualizations to facilitate interpretation of the axes of a (2D) projection, to judge the variation of the data \cite{Kim2016, Vatavu2014Heat} or the relative importance of the data attributes along an axis \cite{Kwon2016}. \citet{Liu2019} used a cartographic approach to compare and analyze learned embedding spaces. In this work, we adapt similar visualization concepts with the goal to create an interpretable gesture space.

To the best of our knowledge, \textit{\system} is the first tool to use a latent variable model to analyze sensor-based motion data in the context of gesture elicitation studies. We combine interactive k-means clustering, automatic metric computation, a new visualization, and analysis concepts to provide an integrated platform.

\section{\system{} Concept}\label{sec:concepts}
We introduce a structured analysis approach based on a learned 2D gesture map, as realised in \textit{\system}. We motivate the conceptual features via related work as summarized in Table \ref{table:design_goals} and elaborate on them in the following sections.

\subsection{Feature Requirements and Overview}\label{sec:feature_requirements_and_overview}

\begin{table}[!t]
    \renewcommand{\arraystretch}{1.5}
    \footnotesize
    \newcolumntype{L}[1]{>{\raggedright\let\newline\\\arraybackslash\hspace{0pt}}m{#1}}
    \centering
    \begin{tabularx}{\minof{\columnwidth}{0.6\textwidth}}{L{2.5cm}L{2.5cm}L{2.5cm}}
        
        \toprule
        \textbf{Challenge / Motivation in the Literature} & \textbf{Visual Analytics Actions} & \textbf{Feature in \textit{\system}} \\
        
        \midrule
        Call for more computational support \cite{Vatavu2016, Villarreal-Narvaez2020, Tsandilas2018}
        &   Model Building, Model Usage
        &   Average Gesture Sequence; Statistical Plot Overlay; Variance Computation \\
             
        Multiple representation for gesture sequences \cite{Villarreal-Narvaez2020, Jang2014}
        &  Visual Mapping
        &  2D Path; 3D Skeleton \\
            
        Comparisons across participants, sessions and trials \cite{Vatavu2016, Jang2014}
        & Visualization Manipulation
        & Selective Filtering; Gesture Highlighting \\

        Visual support for temporal dimension \cite{Jang2014, Nebeling2015}
        & Visual Mapping
        & 2D/3D Animation \\

        Unfamiliarity with Gesture Design Space \cite{Chen2018, Dim2016}
        & Visual Mapping, Model usage, Model-vis
        & Interactive Gesture Map \\

        Processing large data sets \cite{Jang2014, Nebeling2015, Ali2018, Ali2019}
        &   Model Building, Model Usage
        &   Interactive Clustering; Cluster Reassignment \\

        Share and Save Analysis \cite{Magrofuoco2019, Nebeling2015, Jang2014}
        &   N/A
        &   Export Analysis \\
            
        \bottomrule

    \end{tabularx}
     \caption{Main analysis components in \textit{\system} with the challenge and related work that motivated this feature and a reference to the supported action within the \textit{Knowledge Generation Model for Visual Analytics}~\cite{Sacha2014}.}
    \label{table:design_goals}
\end{table}

The features in \textit{\system} were informed by close examination of the literature on gesture elicitation and related concepts and tools: We collected features 1) proposed in related work,  2) motivated in calls for further improvements, and 3) explicitly requested from future work. In addition, we included further ideas. Table \ref{table:design_goals} shows an overview of the relation to related work. The following paragraphs further introduce and motivate the features.

\subsubsection{3D Skeleton View  (Figure~\ref{fig:teaser}b~\textcircled{\small{2}})}
Related tools \cite{Nebeling2015, Jang2014} show a 3D skeleton view with animation. \textit{\system} also offers this, to afford easy examination of a recorded gesture.

\subsubsection{2D Map View  (Figure~\ref{fig:teaser}b~\textcircled{\small{1}})}

\textit{\system} is fundamentally motivated by providing researchers with a visual overview of the elicited gesture space. %

Furthermore, some researchers indicated that participants may struggle to propose gestures, if they are unfamiliar with the gesture design space \cite{Dim2016, Chen2018, Silpasuwanchai2015}. They therefore modified elicitation such that people could choose from a predefined list of gesture proposals.

\textit{\system} addresses these needs as its 2D map shows observed gesture proposals and gives an idea of past behavior. While we focus on researchers as users of this map in this paper, it could also be shown to participants as we described in Section~\ref{sec:opportunities}.

\subsubsection{2D Map Overlays  (Figure~\ref{fig:teaser}b \textcircled{\small{1}}, Figure~\ref{fig:teaser}c)}
Prior work has extensively used scatter plots to analyze machine learned representations \cite{Smilkov2016, Liu2019}. Our map view affords different plots on top of it, such as:
\begin{itemize}
    \item Scatter plots (point = body posture; Figure~\ref{fig:teaser}b \textcircled{\small{1}})
    \item Drawing paths (path = gesture; Figure~\ref{fig:multiple_trials})
    \item Densities (e.g. where in the space are postures and gestures located? Figure~\ref{fig:teaser}c)
\end{itemize}

\subsubsection{Linked Views of Postures}
\citet{Villarreal-Narvaez2020} called for future work to include multiple representations of gestures. \textit{\system} realises this by linking the 2D map and the 3D skeleton. Concretely, the 3D skeleton view updates while the user moves the cursor over the 2D map  to present the posture at that point in the gesture space.

\subsubsection{Linked Animations of Gestures}
Complementary to the feature for postures, \textit{\system} accounts for the temporal nature of gesture data~\cite{Jang2014, Nebeling2015} by offering linked animations of gesture paths (point moving on the path) and 3D skeletons (skeleton moving). %

\subsubsection{Gesture Clustering}
As larger data sets are expected in the future~\cite{Jang2014, Nebeling2015, Ali2018}, we also provide an interactive clustering method to reduce manual workload for identifying similar gesture (sub)groups. 

\subsubsection{Sharing Results}
Motivated by such interests in related work~\cite{Magrofuoco2019, Nebeling2015, Jang2014}, we include an export functionality to easily share analyses with other researchers.

\subsection{The Learned 2D Gesture Map}

Here, we describe the map concept in more detail.

\subsubsection{Core Visualization Concept}
Following a cartographic approach \cite{Skupin2002}, and in line with 2D projections in visual analytics (e.g.~\cite{Wenskovitch2020, Kim2016}), we use a map metaphor to visually guide analysts through the elicited gesture space.
This gesture map is a 2D plot with a grid of representative body poses shown as small human skeletons. These ``pose landmarks'' give an overview of the poses in the corresponding rectangular map region (Figure~\ref{fig:teaser}b~\textcircled{\small{1}}). The map itself is  continuous, that is, each 2D point represents a pose. Thus, since gestures are sequences of poses, they are paths connecting multiple points on the map. In this way, the gesture map combines a line plot's simplicity with the structural expressiveness of a small-multiples visualization \cite{Jang2014}.

\subsubsection{Learning a Gesture Map}
The two dimensions of the map do not have a direct predefined meaning yet emerge from elicited data.
Formally, let the set of all $N$ individual gesture poses in the dataset be denoted by $\mathbf{G}~=~\{g_i ~|~i=~[1,\dots,N]\}$, $g_i \in \mathbb{R}^{D}$ where $D$ is the dimensionality of the raw sensor data \changenote{(in our case D=20). A gesture sequence which consists of $T$ gesture poses can be viewed as an ordered tuple of size $T$ i.e., $\mathbf{g} = (g_1, \dots, g_T)$.}

\begin{enumerate}
    \item To reduce the dimensions of the raw sensor data, we use an encoder $f_{Encoder}:\mathbb{R} ^{D} \to \mathbb{R}^{2}$ to embed every gesture pose $g_i \in \mathbf{G}$ into a latent space code $z_i \in \mathbb{R}^2$. \changenote{These latent codes represent a pose using only two learned features.}  %
    \item \changenote{The raw and high-dimensional gesture sequence $\mathbf{g}$ is then embedded as a two-dimensional path $\mathbf{z} = (z_1, \dots, z_T)$ in the latent space.}
    \item To create the grid of gesture poses in the background, we compute an evenly spaced grid $\mathbf{M} \in \mathbb{R}^{m \times m \times 2}$ of $m$ rows and columns over a visible region in the latent space. \changenote{For example, if the embedded gesture poses (latent codes) range from -4 to 4 in both x and y dimension, we would linearly sample a number of points within this square region.}
    \item \changenote{Using the decoder model we can decode arbitrary 2D map points into a full pose, i.e. $\text{f}_{Decoder}:\mathbb{R}^{2} \to \mathbb{R} ^{D}$}
\end{enumerate}

In this paper we use a Variational Autoencoder (VAE). In general, layout and quality of the space (e.g. smoothness), and of pose decoding, depend on the model, and we reflect on this in our discussion. 

\subsection{Map Interaction Concepts}\label{sec:interaction_concepts}
Here we describe how users can interact with the map.

\subsubsection{Pan and Zoom}
The map supports pan and zoom and accordingly recomputes the grid of landmarks (small skeletons). \changenote{This feature helps to adjust the viewport to support exploration of data-dense areas, and deal with the fact that landmark representations are discrete indicators for the continuous space.}

\subsubsection{Examining Poses}
Scatter or density plots can be projected onto the map (e.g. Figure~\ref{fig:teaser}b~\textcircled{\small{1}} and Figure~\ref{fig:teaser}c). Using ``details on demand'', users can \textit{hover} over points to see the corresponding pose skeleton (Figure\ref{fig:teaser}b~\textcircled{\small{2}}), and referent, participant and trial number in the detail view (Figure\ref{fig:teaser}b~\textcircled{\small{6}}). \changenote{The scatterplot may help researchers to detect outlier body poses, while the density plot reveals regions with recorded data.}

\subsubsection{Examining Gestures}
For further inspection, one or more gestures can be selected (e.g. Figure~\ref{fig:multiple_trials}) from a referent's list of gesture proposals (Figure~\ref{fig:teaser}~\textcircled{\small{3}}). \changenote{This allows researchers to view details on-demand e.g. to reduce the risk of information overload.}

\subsubsection{Examining Unseen Poses}
A fundamentally new capability of \textit{\system} is that \textit{unseen} poses or gestures (i.e. not proposed by participants) can be simulated by decoding arbitrary 2D points in the learned space. In our prototype users can thus hover over the map to visualize 3D skeletons for any cursor location. \changenote{Analysts can examine if empty regions are anatomically not feasible (cf. \ref{dis:unseen}) or if people did not show such behaviour. This might be useful to adjust elicitation setup/instructions, for example to prompt people to also cover a previously empty part of the map.}

\subsection{Analysis Concepts Using the Gesture Map}
Exploratory analysis seeks to uncover structural patterns in the dataset, identify anomalies, and single-out outliers \cite{Tukey1977}. We thus conceptualized the gesture map to enable researchers to seamlessly cycle between the detection of new observations and the assessment of supporting evidence. The analysis concept is structured further by differentiating between global observations and local observations. The former targets questions that may span multiple referents or the entire dataset, while the latter focuses on a few gestures to identify specific behavioral idiosyncrasies.

\subsubsection{Global Observations}
The first of many analysis steps often involves developing an overview of the data to understand its underlying properties: Researchers here often use statistical plots to summarize the data and to identify broad patterns. 

\paragraph{Developing an Overview of the Gesture Space}
\textit{\system} supports this as well: For instance, Figure \ref{fig:scatter_plot} depicts a scatter plot projected on the gesture map. Scatter points on top of the pose grid enable researchers to quickly identify which general poses were observed in the data. Each scatter point corresponds to a pose from the dataset, whereas empty patches in the gesture map may indicate behavior that has not been observed (e.g. poses/gestures not proposed by participants during elicitation).

\paragraph{Spotting Clusters and Outliers}
Scatter points may visually cluster near gesture poses that are characteristic for a particular referent. These clusters can help researchers to form a mental model of the main poses that are characteristic for a group of gesture sequences. It might also be interesting to analyze outlier behavior which can be detected by examining scatter points that lie far from these clusters.

\paragraph{Comparing Referents and Regions}
Additionally, color codes facilitate the comparison of behavior across different referents. For example, it might be interesting to identify which referents share behavior and which are distinctive. Regions in the gesture map that contain multiple embedded data points from different referents may indicate that this region encodes shared generic behavior. 

\paragraph{Judging Densities and Overlap of Referents}
Scatter plots may contain too much detail and clutter the visualization. Density plots then offer a visualization of the most frequent gesture poses. Researchers can use it to detect overlapping or distinctive behavior across different referents. For example, these observations can inform researchers interested in building gesture recognizers in judging the difficulty of separating gestures for the various referents.

\subsubsection{Local Observations}
The key local observation in elicitation data is to \textit{examine individual gesture proposals}. \textit{\system} also supports such analysis, as outlined here:

After the initial data exploration it is often necessary to find concrete example for detected patterns. For example, in the elicitation context, we might be interested in comparing the behavior across different participants and experimental trials. 

The gesture map can serve as a common visual basis for such investigations: By projecting multiple gestures onto the map, researchers can evaluate each participant's behavior individually. The trajectory of the embedded gesture paths can inform them on specific behavioral characteristics. For example, a participant's movement can be subtle, in which case the embedded gesture path is simple in shape and typically spans a small region in the gesture map. In contrast, a complex gesture may be represented as an intricate path that may meander across the map. 

Thus, by comparing multiple embedded gesture paths researchers can visually assess gestures as similar or not. Considering research interests in the elicitation context from the literature, for example, this might support researchers to examine if a participant can remember and repeat the same gesture proposal across multiple trials \cite{Nacenta2013}, or if behavior was influenced by a priming effect \cite{Cafaro2018}.

\section{Consensus and Clustering with DBA}\label{sec:computational_methods}
We introduce the concept of an \textit{average gesture sequence} as a new computational capability in the context of gesture elicitation. This has three practical values, which complement our tool: 
\begin{enumerate}
    \item \textit{Descriptive:} The average gesture can serve as a single, visual proxy for a group of gestures, which opens up new visualization opportunities (e.g. showing and comparing referents as average paths on our gesture map). 
    \item \textit{Evaluative:} It enables a new measure of consensus/variability among gesture proposals for a referent. This measure aligns well with other statistical notions of variability: Consensus is assessed via the variance of actual gestures around the average gesture.
    \item \textit{Explorative:} The average gesture enables clustering methods that require averaging (e.g. here: k-means), supporting the automated detection of groups of gestures in a dataset (e.g. in ``open elicitation'' without referents, cf.~\cite{Villarreal-Narvaez2020}). 
\end{enumerate}

We next describe the technical approach in more detail.

\subsection{Computing an Average Gesture with DBA}
We employ the DTW Barycenter Averaging (DBA) algorithm by \citet{Petitjean2011} to compute an average gesture: Intuitively, this algorithm first aligns an initial sequence with every sequence in the set of gesture proposals, before computing a centroid (barycenter) for each aligned coordinate. For further technical details we refer the reader to the related work~\cite{Petitjean2011}.

\subsection{Consensus as Variation Around Barycenter}\label{sec:consensus_dba}
\citet{Vatavu2019} were the first to propose a data-driven consensus measure that does not rely on human judgement of gesture similarity. To achieve this, they employed Dynamic Time Warping (DTW) distance computations to define a consensus measure: They considered two gesture sequences $\mathbf{g}_A$ and $\mathbf{g}_B$ as similar if the DTW distance was below a threshold $\Delta_{DTW}(\mathbf{g}_A, \mathbf{g}_B) \leq \tau$. To determine consensus for a referent they calculated the \textit{pairwise distances} across all gesture proposals for this referent. Finally, to report a measure independent of the threshold value $\tau$, they used a \textit{logistic regression model} to determine the consensus for a range of normalized threshold values and reported the \textit{growth rate} as an indication of the overall consensus.

This work motivates us to further explore data-driven measures of consensus: We follow a similar approach, but instead of regressing on the DTW distance values, and relying on pairwise comparisons, we directly compute an average sequence from all gesture proposals in a referent group, using DBA.

We then measure the DTW distance of every gesture proposal $g^*$ for a referent $R$ to the computed average gesture (i.e. barycenter) $\mathbf{g}_{DBA}$ for $R$. Finally, we  report the variance of these DTW distances as a measure of consensus. %
Formally, this is noted as:
\begin{equation}
    \textit{VAR}_{R} = \frac{\sum_{\mathbf{g^*}}^{\mathbf{G}_{R}} \left(\Delta_{DTW}\left(\mathbf{g^*}, \mathbf{g}_{DBA}\right)\right)^2}{|\mathbf{G}_{R}|}
\end{equation}
$\mathbf{G}_{R}$ denotes the set of all gestures elicited for referent $R$.
Intuitively, for example, a high value $\textit{VAR}_{R}$ may inform an analyst that referent $R$ contains quite varied gesture proposals (i.e. low consensus). 

The gesture variance integrates well with \textit{\system}'s visualization concept because this already displays the involved average gestures as visual elements. Moreover, this approach yields a one-number summary without a logistic regression model on top. Overall, we see this approach as an additional measure, not a replacement of others: As a flexible tool, \textit{\system} can be extended to additionally display further such measures (e.g. the one by \citet{Vatavu2019}) to support researchers with the analysis.

\subsection{Clustering Gestures with DBA \& K-Means}
Being able to compute an average gesture enables the use of clustering methods that require average computations. Here, we use k-means in particular. %
The idea of clustering gesture elicitation data is motivated by two aspects:

\begin{enumerate}
    \item \textit{Exploration:} For example, in ``open elicitation'' \cite{Villarreal-Narvaez2020} or settings where referents are not predefined, such as in the work by \citet{Williamson2012}, a clustering may proved a valuable reference point to identify novel behavior. 
    \item \textit{Annotation:} Clustering may also be used to help kickstart (manual) annotation in cases where explicit groupings of proposals are desired (e.g. for agreement measures \cite{Wobbrock2009}).
\end{enumerate}

Considering the literature, \citet{Jang2014} used an interactive hierarchical clustering approach with complete-linkage. In contrast, we experimented with the k-means algorithm, using DBA to calculate the centroids. We motivate this choice by interpretability of the resulting centroids, versus the abstract representations in the hierarchical approach: In particular, the centroids (i.e. average/barycenter gestures) are more compatible with our 2D gesture map, on which they could be displayed as paths. In contrast, a hierarchical treemap does not directly fit the map metaphor well.

\section{Implementation of \system}\label{sec:tool}

We implemented \textit{\system} as an analysis tool that integrates the described concepts of both the interactive gesture map (Section~\ref{sec:concepts}) and the DBA-based computations (Section~\ref{sec:computational_methods}). Here we describe the key implementation aspects.

\subsection{User Interface and Functionality}
Figure~\ref{fig:teaser} shows the UI; the following sections refer to the numbers in the figure. Overall, we implemented all UI views and interactions conceptually described in Section~\ref{sec:concepts}.

\subsubsection{Gesture Map Figure~\ref{fig:teaser}b \textcircled{\small{1}}}
Researchers can zoom, pan, and hover over the gesture map, and overlay a scatter plot or a density plot (e.g. Figure~\ref{fig:teaser}c) to explore individual or multiple gesture poses.%

\subsubsection{Experiment View Figure~\ref{fig:teaser}b~\textcircled{\small{3}}}
This view lists all referents and gesture proposals in a compact way as numbers for quick reference and selection. When hovering over an element, the corresponding gesture path is shown on the map for a moment.

\subsubsection{3D Skeleton View Figure~\ref{fig:teaser}b~\textcircled{\small{2}}, Figure~\ref{fig:teaser}b~\textcircled{\small{4}}}
This view either shows the raw skeleton recording or a reconstructed skeleton. If researchers animate a gesture, it is simultaneously animated in this view and on the map. The progress of the animation can be controlled via a play/pause button and slider.

\subsubsection{Statistics View Figure~\ref{fig:teaser}b~\textcircled{\small{5}}}
This view shows different metrics, namely variances around the average gesture sequence per selected referent (Section~\ref{sec:consensus_dba}), the distributions of DTW distances of proposals to their average gesture sequence, and nearest neighbor distances for a selected gesture.

\subsubsection{Cluster View Figure~\ref{fig:teaser}b~\textcircled{\small{6}}}
This dialog is unfolded with a button in Figure~\ref{fig:teaser}b~\textcircled{\small{3}} and lets users interactively cluster gesture proposals for a referent. Centroids can be animated and once the clusters have been computed, users can toggle all gesture proposals that were assigned to a centroid.

\subsection{Architecture}
\changenote{We used a server-client architecture. The frontend and backend modules communicate through a REST API through which the data is transmitted as a JSON formatted string. The frontend was implemented with NodeJS~\citep{NodeJs} and React~\citep{React}. For plotting, we use the PlotlyJs library~\cite{Plotly}.
For the backend we used the Flask framework~\cite{Flask} and Pandas~\cite{Pandas} to handle the data transformations and queries. We cached expensive computations such as the computed average sequences and distances matrices on MongoDB~\cite{MongoDB} to . PyTorch~\cite{Paszke2017} was used to develop the embedding model.}

\section{Experiments}\label{sec:experiments}
Ledo et al.~\cite{Ledo2018} identified four evaluation strategies for toolkit contributions. We follow their perspective to evaluate \textit{\system}, combining two such strategies: 
First, here we follow the \textit{Demonstration} strategy and provide a detailed analysis of examples on elicitation data from related work.
Second, Section~\ref{sec:user_study} follows the \textit{Usage} strategy and reports on a user study with HCI researchers. %

\subsection{Datasets}
We consider four existing datasets: One explicit gesture elicitiation study by \citet{Vatavu2019}, plus three datasets collected for gesture recognition systems~\cite{Aloba2018, Fothergill2012, Chen2015}. 
We first focus on the dataset by \citet{Vatavu2019} that consists of 1312 full body gestures elicited from children aged 3-6, recorded with a Kinect sensor. %
For preprocessing, we followed the original authors~\cite{Vatavu2019} but left out the resampling step. %

\subsection{Model Training}
We used a Variational Autoencoder (VAE)~\cite{Doersch2016} to embed the data as a 2D gesture map. The VAE here serves as an exemplar of a model with both powerful (non-linear) encoding and decoding capabilities. We reflect on other possible choices in our discussion.

We trained the VAE on the poses (frames) of the mentioned dataset~\cite{Vatavu2019} which has 60 dimensions (20 body joints $\times$ $x,y,z$). %
We adapted the architecture from \citet{Spurr2018} (i.e. 4 hidden layers for both encoder and decoder) and used a 2D bottleneck layer.  %
In line with Fu et al. \cite{Fu2019}, we used a weight term to modulate the mix of KL-loss and reconstruction loss in early training. We trained for 2000 epochs with Adam~\cite{Kingma2014} (lr=$3e^{-5}$).  %

We experimented with different numbers of hidden neurons $h$: %
Overall, reconstruction loss decreases for larger models, regularized by the KL-loss, leading to diminishing returns and a decision for $h=512$ here. For full details, we provide the training scripts and model comparisons on the project website.

\subsection{Global Observations}\label{sec:experiments_global}
\begin{figure}[t!]
    \centering
    \includegraphics[width=\minof{\columnwidth}{0.5\textwidth}]{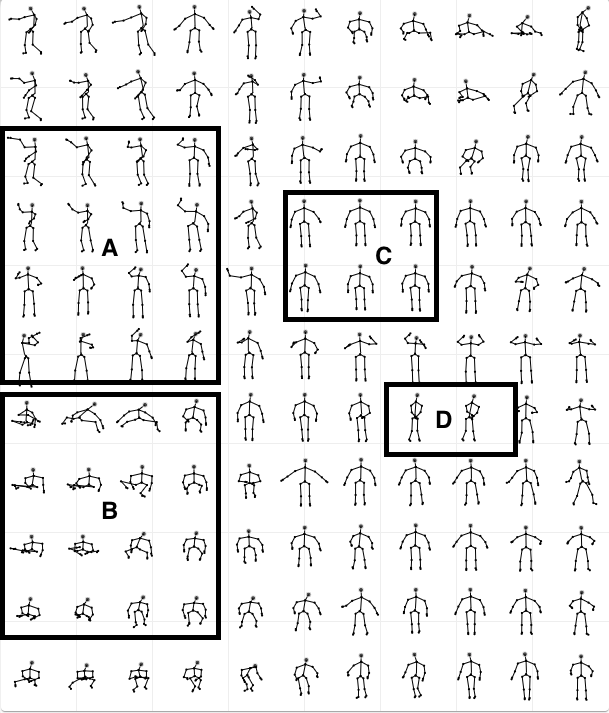}
    \caption{Gesture map for the dataset by \citet{Vatavu2019}. Pose landmarks represent poses in that part of the learned gesture space. Marked areas are referenced in Section~\ref{sec:experiments_global}.}
    \label{fig:skeleton-grid}
    \Description[]{A grid of body poses representing the space of gesture poses. Four boxes highlight different areas in this visualization and named A through D. The concrete description is given in the text.}
\end{figure}

Here demonstrate the use of \textit{\system} in a walkthrough of an explorative analysis: %
Examining the gesture map, the center (Figure~\ref{fig:skeleton-grid}C) reveals start/end poses (standing upright, arms at rest). %
We further see, for example, sitting (Figure~\ref{fig:skeleton-grid}B), clapping (Figure~\ref{fig:skeleton-grid}D), and raising an arm (Figure~\ref{fig:skeleton-grid}A). 
Thus, the map reveals the space of poses elicited by \citet{Vatavu2019} at a glance: For example, their referents included \textit{crouch}, \textit{draw a flower}, \textit{draw a circle}, \textit{draw a square}, \textit{applaud} or \textit{raise your hands}, which all match the poses in our map.

\begin{figure}[t!]
    \centering
    \includegraphics[width=\minof{\columnwidth}{0.5\textwidth}]{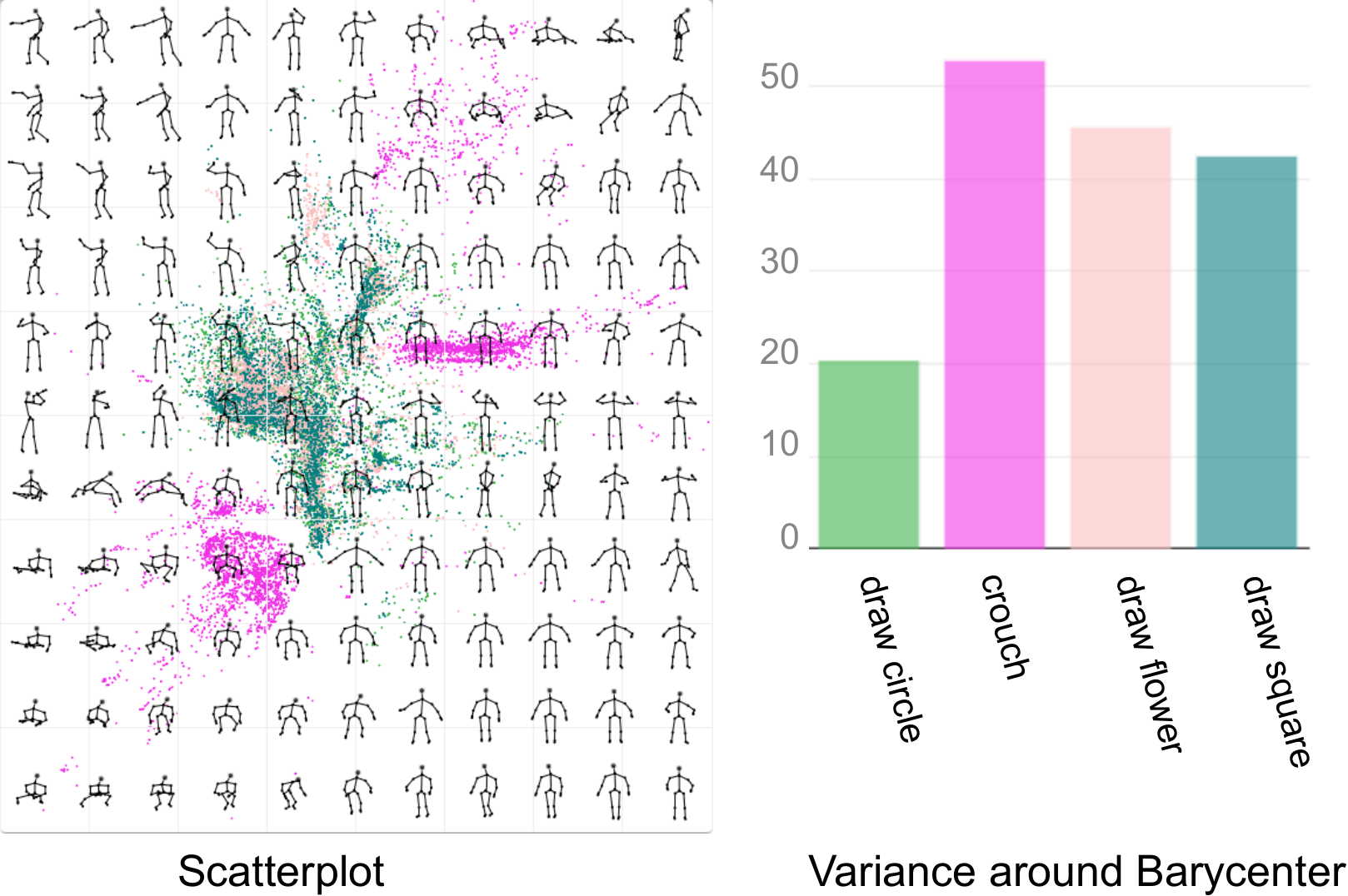}
    \caption{Left: Scatter plots show gesture poses for four referents elicitet by \citet{Vatavu2019} (\textit{crouch, draw circle, draw flower, draw square)}. Right: Variances of the gestures' DTW distances to their average gesture sequence.
    }
    \label{fig:scatter_plot}
    \Description[]{Two plots are shown side-by-side. The plots highlight that the variance of the gestures around their average sequence differ between different referent groups. In descending order the variances are largest for crouch, draw flower draw square, and finally, draw circle.}
\end{figure}

Using overlays in \textit{\system}, we can identify similarity and differences between gestures across referents: For example, Figure \ref{fig:scatter_plot} (left) shows that \textit{crouch, draw circle, draw flower, draw square} share common behavior; their scatter points largely overlap in the region that encodes ``raised arm'' behavior. In contrast, for instance, gestures proposed for \textit{crouch} cover a different region (pink). 

The variance plot in \textit{\system} (Figure \ref{fig:scatter_plot} right) indicates that proposals for  \textit{crouch} and \textit{draw flower} vary more than for \textit{draw circle} and \textit{draw square}. Potentially, for the children the basic shapes afforded less flexible interpretation than a flower or crouching.

We defined a consensus measure on this variability (Section~\ref{sec:consensus_dba}):
Comparing this variability between all referents, our results largely agree with \citet{Vatavu2019}: In particular, \textit{applaud, fly like a bird} and \textit{hands up} show high consensus while \textit{climb ladder, crouch, turn around} have low consensus.

\subsection{Local Observations}

\begin{figure}[t!]
    \centering
    \includegraphics[height=6cm]{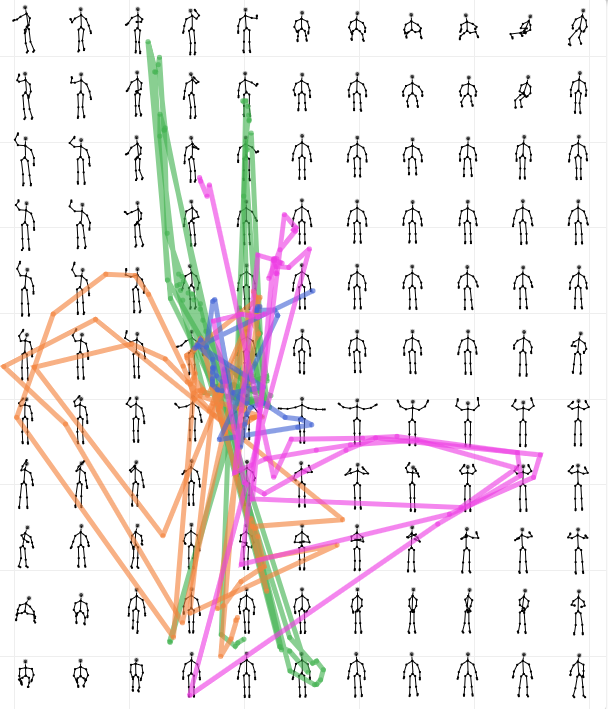}
    \caption{\changenote{Gesture proposals for \textit{throw ball} from four people (different colors). Trials per person are not discernible (same color), yet the colored paths distinctly cover different regions, revealing high consistency per person.}}
    \label{fig:multiple_trials}
    \Description[]{Multiple embedded 2D gesture paths are projected on the gesture map. The gesture sequences correspond to different study participants. Each participant's gestures cover a region of the gesture map that is different from the other participants.}
\end{figure}

Proposals for \textit{crouch} form two main clusters (pink points in Figure \ref{fig:scatter_plot} left), one in the region of starting poses, another in sitting/crouching regions. 
Thus, \textit{\system} visually reveals that people interpreted \textit{crouch} in different ways, matching the high variance (Figure \ref{fig:scatter_plot} right). Examining the map locally, in combination with gesture animations, reveals that some children sat on the floor, some on their heels, some crawled on hands/knees, and others stood with a stooped body posture. Some additionally jumped at the end of their gesture proposals to get back onto their feet.

As another such example, for \textit{throw ball}, behavior can be categorized into four clusters: Most children used their right hand, others used two hands, and some kicked the ball. Only a few used the left hand. As Figure \ref{fig:multiple_trials} shows, the children mostly stuck to their interpretation across multiple repeated trials for that referent, revealing consistency (cf.~\cite{Vatavu2013a}). This is an example for using \textit{\system}'s spatial visualisation of gestures as paths for visual comparison via shape.

\subsection{Interactive Clustering}
\changenote{For a typical elicitation study, such as this one by \citet{Vatavu2019}, it is reasonable to expect clusters induced by the referents. Therefore, to demonstrate our proposed clustering analysis we removed the referent labels and then evaluated if k-means finds clusters that match the original referents.}

Concretely, we ran the clustering with 15 sequences chosen randomly. %
We then inspected the mix of original referents present in the gestures assigned to each found cluster. %
We repeated this ten times and made these observations: 

\begin{itemize}
    \item Our k-means clustering identified those referents with high agreement (e.g. \textit{hands up, crouch, applaud, fly like a bird}).
    \item Gestures for referents with much common behavior appeared as one cluster (e.g. \textit{draw circle, square, flower}). Note that this is not necessarily ``wrong'', since a behaviour ``draw something'' would also have been a plausible referent.
    \item The resting pose was detected as a separate cluster.
    \item Other referents were (clearly) present only in some of the clustering repetitions.
\end{itemize}
Overall, this indicates the potential of automated clustering, for example, when examining data from open elicitation with no given referents. We return to ideas for improvements in our discussion.

In another experiment, we applied clustering to look for patterns within a referent: As mentioned, referents such as \textit{throw ball} and \textit{crouch} contained distinct patterns, revealed on the map. %
Indeed, running k-means revealed some of them: For example, for \textit{throw ball} k-means also detected throwing with the right hand vs using both hands. In contrast, it did not separately find left hand and kicking, presumably since those were proposed only a few times. %

\subsection{Comparison Between Datasets}
Other researchers noted that elicitation findings are spread across multiple venues and need to be consolidated \cite{Villarreal-Narvaez2020}. \textit{\system} supports this as it offers a platform to visualize and analyze multiple studies. We demonstrate this by creating a gesture map using \textit{four} datasets \cite{Aloba2018, Fothergill2012, Vatavu2019, Chen2015a}. 

\begin{figure}[t!]
    \centering
    \includegraphics[width=\minof{\columnwidth}{0.5\textwidth}]{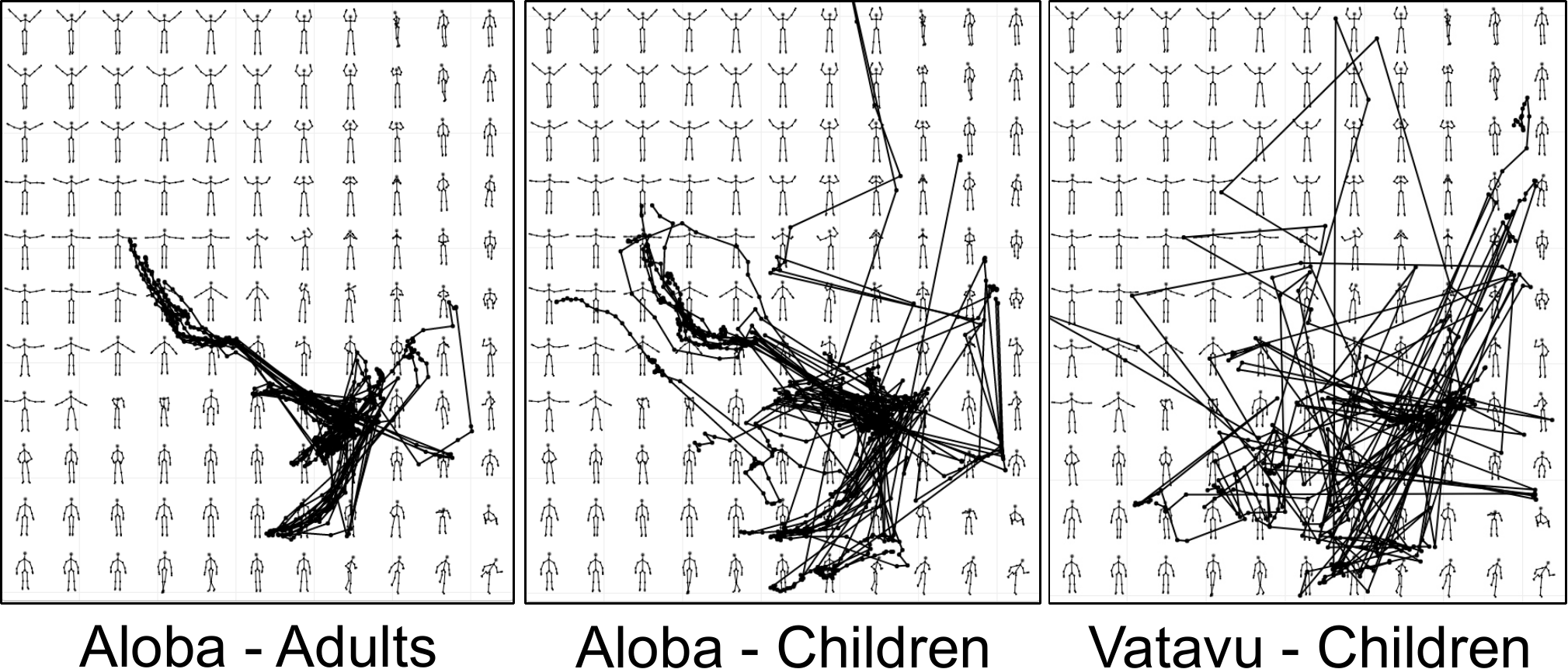}
    \caption{Gesture paths for adults and children for ``jumping'' referents from two studies~\cite{Aloba2018, Jain2016}.}
    \label{fig:adultvschild}
    \Description[]{Three plots are plotted side-by-side. Each plot shows multiple gesture proposals as embedded 2D paths on-top of the gesture map. The gesture sequences are more aligned for adults than for children.}
\end{figure}

To motivate a concrete example, citet{Jain2016} showed that observers can distinguish behavior of children and adults. Figure~\ref{fig:adultvschild} shows all 20 proposals for \textit{jump} from the data by \citet{Aloba2018}, next to the children's proposals from \citet{Vatavu2019}. The gesture paths visit roughly similar main parts of the gesture space, yet the children do not find consensus. Our variance measure also reflects this (Aloba - adults $23.70$, children $44.89$; Vatavu - children $54.94$). %

\begin{figure}[t!]
    \centering
    \includegraphics[height=6cm]{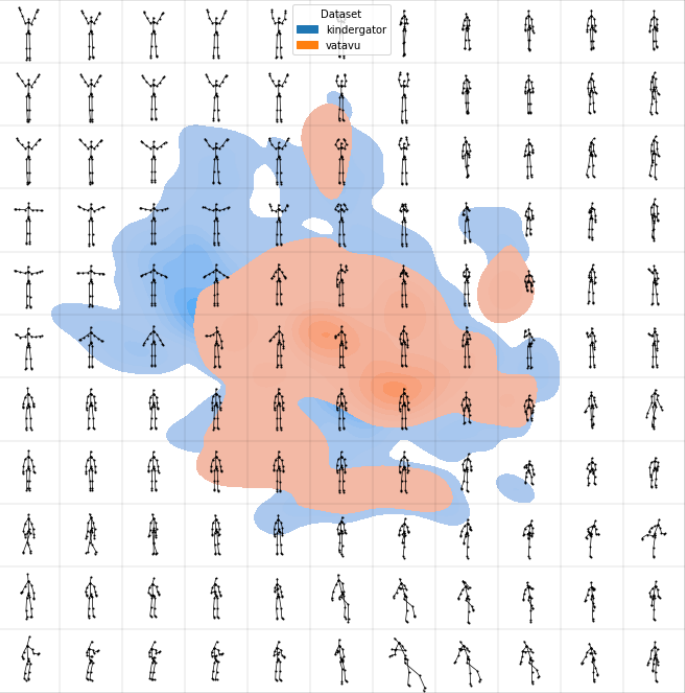}
    \caption{Combined gesture space from \cite{Aloba2018, Fothergill2012, Vatavu2019, Chen2015a}. The density plots projected on this gesture map refer to \cite{Aloba2018} (\textit{blue}) and \cite{Vatavu2019} (\textit{orange}).}
    \label{fig:cmp_kg_vat}
    \Description[]{Two density plots show the distribution of body poses from two datasets. One density plot covers a larger area of the gesture map than the other density plot. The larger plot therefore indicates that the corresponding dataset involves many diverse body poses.}
\end{figure}

As a second example, we compared \textit{behavior diversity} across datasets. Without knowing anything about the referents, Figure~\ref{fig:cmp_kg_vat} already reveals that one dataset~\cite{Aloba2018} (\textit{blue}) covers a larger region than the other~\cite{Vatavu2019} (\textit{orange}). Thus, it seems to contain a more diverse set of body poses. Indeed, this observation can be explained by the longer referent list (58 referents in \cite{Aloba2018} vs 15 in \cite{Vatavu2019}).

\section{User Study}\label{sec:user_study}
To further evaluate \textit{\system}, we recruited eight HCI researchers (7 male, 1 female) from three universities via e-mail for remote think-alound and interview sessions. Six were familiar with gesture elicitation studies, the other two were interested in analysing gesture sensor data. Five were familiar with machine learning. %

\subsection{Procedure}
The interviews lasted 80 minutes %
and were conducted via screen-sharing using Skype/Zoom, with \textit{\system} hosted online such that people could use it on their own computer. We again used the dataset by \citet{Vatavu2019}. %
With people's consent we recorded the interviews.
We encouraged them to think out loud and occasionally asked questions to better understand actions. We took notes and compiled a report from this material. Given the exploratory nature of the interactions and the diversity in people's approaches this was done in an inductive approach, leading to the themes in Section~\ref{sec:findings}.

The interviews had four parts: 1) We introduced \textit{\system} (20 minutes), with a concept presentation, a guided walk through the tool and UI, and opportunities for questions. %
2) In an \textit{exploratory, manual analysis task} people were prompted to use \textit{\system} to identify groups of behaviors in the gesture proposals for two referents. In real use, researchers would conduct such analyses to better understand elicited data.
3) In a more \textit{confirmatory, automatic analysis task} we asked them to build on their gained insights to initialize the clustering algorithm and refine the automatic clustering results. In real use, researchers might export this result, for example, for a report, calculations of agreement, etc.
4) The session concluded with a semi-structured interview of at least ten minutes. Here, we inquired into what people liked/disliked about \textit{\system}, and asked for ideas for improvements and additional features.  %

\subsection{Findings}\label{sec:findings}

\subsubsection{Initial Use}
Upon first use, most people immediately animated a few gestures, saying that this was the most natural and familiar way to view the data 
Since the map visualization was unfamiliar to them, some had initial difficulties to understand the distinction of single poses (points) and entire gestures (paths). These people found the animation particularly important: Seeing the 3D skeleton and the 2D path animated in sync highlighted that a gesture was a path on the map and thus helped them to get familiar with the map concept. %
Summarising their initial experience, one person said: \textit{``Although, the learning curve [...] is steep, once you understand the core concepts, this tool offers a great overview of the entire behavior that is captured in the dataset.''}

\subsubsection{Statistical Plot Overlays}
We asked the researchers to analyze the proposals for \textit{crouch} and \textit{throw ball}. \changenote{Throughout the interview we noticed that all participants preferred the scatter plot over the density plot. When asked why they keep returning to the scatter plot, they said that it provided more detail and that density can also be estimated from scatter points.} They also said that points were visually closer to the data (point=pose).

\subsubsection{Details of the Gesture Map View}
\changenote{When study participants paused their exploration for a longer period, we inquired why that was the case.} Some people noted that they struggled to find a specific pose on the map. \changenote{They suggested to increase the visibility of the poses by showing fewer and larger landmarks.} %
\changenote{Another researcher felt that the map should show more detail so it would be easier to judge differences and transitions of poses.} %
Together, this feedback motivates a changeable grid size (our zoom was implemented to always keep an 11 $\times$ 11 grid).

Some found similar poses encoded in different map regions and noted that these should ideally reside in one area. This is an artefact of dimensionality reduction, as we discuss further in Section~\ref{sec:diss:methodchoice}

\subsubsection{Exploration Strategies}
\changenote{When we asked the participants what the main aspect was that they used to determine interesting behavioral patterns, we observed diverse analysis strategies, but we broadly highlight two main ones:} 

1) \textit{Shape driven analysis:} Some started by skimming through gestures to get an overview of their different path shapes on the map. They stopped to examine gestures in more detail that differed largely from the shapes seen so far. In a sense, they searched for outlier behavior based on the path shapes. These participants noted that the 2D gesture path visualization offers a quick way to spot irregular behavior and that their analysis becomes an active search versus passively watching every gesture individually. 

2) \textit{Position driven analysis:} In contrast, other participants focused entirely on the scatter points as template poses. Using expectations about possible behavior for a gesture proposal (e.g. left vs right hand throwing), they examined scatter points in those map regions that based on the landmark skeletons encoded related poses. %

\subsubsection{Manually Forming Clusters}
\changenote{Regardless of their initial analysis strategy, when asked which feature they would use to group the gestures, people agreed on the path shapes as primary discerning feature (strategy 1).} For the \textit{crouch} referent, everyone distinguished two to three groups of behaviors. For \textit{throw ball}, everyone found at least three (left/right/both handed throwing). Some also found the kicking behavior as described in Section~\ref{sec:experiments}. Overall, the researchers felt comfortable with grouping the proposals based on the path shapes. However, there were some complex paths (e.g. crossing over many poses on the map) that people were unable to assign to a group. One person suggested to create an outlier group for these.%

\subsubsection{Interactive Clustering}
We asked people to use the interactive clustering tool based on their observations in the first task. 

\changenote{Next, they were asked to initialize the clustering algorithm using their knowledge from the previous task. Now, all participants specifically searched for individual gesture proposals as templates (strategy 2) and used those to initialize the algorithm.}

However, the resulting computed centroids often deviated from people's expectations, %
and thus did not immediately make sense to them. 
One user noted that one still has to inspect all gesture proposals in order to choose suitable initialisations for the k-means algorithm. 
On the positive side, the researchers liked the refinement step, where they could reassign proposals to another cluster. 
These reassignments, however, were not yet considered when rerunning the clustering algorithm in the current implementation.

Overall, after being asked to give a final verdict over the interactive clustering feature, all deemed it important. However, they noted that it should be more accurate and manually refined assignments need to be respected when rerunning the clustering algorithm, thus enabling iterative, interactive use. Technically, this can be readily implemented by initialising k-means with the current (refined) assignments.  %

\section{Discussion}

\subsection{Extending the Gesture Elicitation Toolbox}
\textit{\system} builds on and extends functionalities of previous tools for gesture elicitation: It combines 1) gesture modeling and visualization, 2) automatic computation of elicitation metrics, and 3) interactive clustering to provide an integrated analysis platform. 

Seeing this and related work as a ``toolbox'', researchers may now consider various options:
For example, AGATE 2.0~\cite{Vatavu2016} is a highly specialized tool to compute agreement, which assumes a given labeled dataset. \textit{\system} could be used to label data and export it for analysis in tools like this.

Alternatively, \citet{Ali2018} proposed a crowd platform for annotation, yet without computational support for the workers, such as alternative gesture representations or similarity measures. Such support as shown in \textit{\system} could be combined with a crowd approach in the future. \textit{\system} is already implemented as a web-based tool, rendering it flexible and open to such integration.

Looking ahead, new cloud elicitation tools \cite{Magrofuoco2019, Ali2019} yield large datasets. \textit{\system}'s concepts support handling large data, visually summarised and explored via our map view. %

Finally, the ``toolbox'' in the literature includes several formalized agreement measures~\cite{Wobbrock2009, Vatavu2012}. These could be used also with our interactive clustering, for example, by plugging in the cluster cardinalities instead of subjective gesture group counts. %

\subsection{Reflection on Model \& Clustering Choices}\label{sec:diss:methodchoice}
Here, we  highlight model and clustering aspects to consider. %

\subsubsection{Smoothness of the Latent Space}
A smooth latent space facilitates suitable visualization by reducing ``jumps'' in gesture paths. These occur due to recording issues (e.g. sensor occlusion in some frames) or when subsequent poses are embedded far apart in the 2D space. While some models address this (e.g. we used a VAE instead of AE), there is no universal ``natural'' 2D layout of body poses and some artifacts are likely to exist for most models and datasets. Besides technical model improvements, visualization concepts could be explored to address this as well (e.g. visually mark ``jumps'' along the gesture path). %

\subsubsection{Cluster Approaches}\label{sec:diss:clustering}
A difficulty with k-means is setting the number of clusters. %
As an example strategy, to detect the subgroup behavior for the \textit{throw ball} referent, we quickly skimmed through the gestures using the map and visually identified rough patterns. We then chose $k$ correspondingly. %
\changenote{We chose k-means, because it readily integrates with the gesture map and the "variance around mean gesture" that we introduced in section \ref{sec:consensus_dba}. Color coding the cluster results can be done quickly. \citet{Jang2014} proposed to use interactive hierarchical clustering. Integrating such a tree-like layout into the gesture map adds complexity and might be material for future endeavours. We can imagine that average gestures calculated with the DBA-algorithm can be used to visualize the non-leaf nodes in the hierarchical tree. In addition, interactive hierarchical clustering would eliminate the need for choosing the number of clusters beforehand.}

\subsubsection{Feature Representation}

\changenote{Hand-engineered features \cite{Jang2014, Aloba2020, Vatavu2017} may help with the interpretation, however, they may be specific to a sensor and interaction setup. As an exploratory tool, \textit{\systems} learned space is applicable to new and changing setups, without developing hand-engineered features first. Furthermore, our learned representation supports gesture simulation useful to examine regions of the behavior space that were not covered by participants.}

\subsection{Opportunities for Research \& Applications}\label{sec:opportunities}
Here we outline further ideas enabled or supported by \textit{\system}.

\subsubsection{Supporting Meta-Analysis and Consolidation}
\textit{\system} empowers researchers to compare data across studies (cf. Section~\ref{sec:experiments}). 
As a community, we could consolidate our findings in a meta-map of many studies, as a sensor data-driven complement to literature surveys~\cite{Villarreal-Narvaez2020}. For instance, such a map might reveal which gestures and poses are most common or intensely studied. Separate maps could also compare gesture spaces for different contexts, devices, etc., for example, to better understand the influences of such factors.

\subsubsection{Enabling Map-based Gesture Authoring}
\textit{\system} could be extended to define new gestures: For example, users could \textit{draw a gesture} as a path on the map. Since the underlying latent variable model can simulate new behavior (decoding), such a drawn path implicitly defines a pose sequence that could be exported as a template-based gesture recogniser.
As an alternative to drawing, users could \textit{demonstrate} the gesture in front of the sensor, with a ``cursor'' moving on the map live.
Users could also \textit{select} recorded gestures on the map, labelled manually or with help from our clustering tool, to train a classifier. Such a recognizer then also could be used in other tools to support sensor feed annotation (e.g.~\cite{Nebeling2015}).

\subsubsection{Enabling Analysis of Unseen Behavior}\label{dis:unseen}
So far, elicitation has focused on observed gestures, yet it might also be relevant to examine why behavior was not observed. 
\textit{\system} enables this: Researchers can explore map areas \textit{without} data, which may reveal unlikely behavior, or indicate issues %
with interaction (e.g. anatomically difficult or tiring gestures) or the sensor (e.g. gestures leading to self-occlusion of body parts).
In this way, \textit{\system} supports the diagnosis of challenges and limitations in the joint user-sensor space of an interactive system (cf.~\cite{Williamson2012}).

\subsubsection{Supporting Live Exploration and Monitoring}
\textit{\system} could be extended to more than post-hoc analysis: For example, we could embed live sensor data and continuously update the underlying mode. 
This live embedding provides  a monitoring tool, for example, for participants to see their currently performed gesture (e.g. shown as a ``cursor''/point on the map), possibly to nudge them towards exploring new regions of the behavior space (cf. \cite{Williamson2012}). 
One could also predefine a gesture path to monitor live performances and to judge deviation from this ``template'', possibly to learn/teach a movement sequence. %
Related, gesture sets are mostly presented as drawings and videos today~\cite{McAweeney2018}. Instead, \textit{\system} could be used to show gestures to users, allowing them to \textit{reenact} and \textit{explore} them with live monitoring via the map. %

\section{Conclusion}
As our key contribution, we presented a set of visualization and analysis concepts for gesture elicitation data and a tool that implements them: \textit{\system} is the first visual analytics tool for gesture elicitation which directly visualises the space of gestures, using a learned 2D embedding. It further leverages the computation of average gestures to enable researchers to 1) represent gesture groups with one gesture; 2) assess consensus as variance around this average gesture; and 3) cluster gestures automatically.

Expert users especially liked the visual expressiveness of \textit{\system}, as it quickly summarizes the underlying dataset. The extensibility of \textit{\system} further encourages future work to employ machine learning as a tool for analysis of human behavior. With this work, we contribute to the vision of more widespread use of applicable computational methods in HCI, also to support more extensive and cost-efficient large-scale, data-driven HCI work. Given the proliferation of crowd platforms to collect large datasets, we expect computational methods and visual analytics as proposed here to become indispensable tools for many future HCI studies.

\textit{\system} and further materials are available on the project website:
\url{https://osf.io/dzn5g/}

\begin{acks}
This project is funded by the Bavarian State Ministry of Science and the Arts and coordinated by the Bavarian Research Institute for Digital Transformation (bidt).
\end{acks}

\bibliographystyle{ACM-Reference-Format}
\bibliography{bibliography}

\end{document}